\documentclass[preprint2]{aastex}
\usepackage{graphicx}
\usepackage{txfonts}
\usepackage{natbib}

\shorttitle{Slow Radiation-Driven Stellar Winds}
\shortauthors{Cur\'e et al.}

\begin{document}
\title{Slow Radiation-Driven Wind Solutions of A-Type Supergiants}
\author{M. Cur\'e} 
\affil{Departamento de F\'{\i}sica y Astronom\'{\i}a, Facultad de Ciencias, Universidad de Valpara\'{\i}so\\
Av. Gran Breta\~na 1111, Casilla 5030, Valpara\'{\i}so, Chile}
\email{michel.cure@uv.cl}
\and
\author{L. Cidale\altaffilmark{1}}
 \affil{Departamento de Espectroscop\'{\i}a, Facultad de Ciencias Astron\'omicas y Geof\'{\i}sicas, 
 Universidad Nacional de La Plata (UNLP), 
and\\ Instituto de Astrof\'{\i}sica La Plata, CCT La Plata, CONICET-UNLP\\
Paseo del Bosque S/N, 1900 La Plata, Argentina}
\and
\author{A. Granada}
\affil{Departamento de Espectroscop\'{\i}a, Facultad de Ciencias Astron\'omicas y Geof\'{\i}sicas, 
Universidad Nacional de La Plata (UNLP), and
Instituto de Astrof\'{\i}sica La Plata, CCT La Plata, CONICET-UNLP,\\ Paseo del Bosque S/N, 1900 La Plata, Argentina
\and
Observatoire Astronomique de l'Universit\'e de Gen\`eve\\ 51, Chemin des Maillettes, CH-1290, Sauverny, Suisse.}

\altaffiltext{1}{Member of the Carrera del Investigador Cient\'{\i}fico, CONICET, Argentina}


 \begin{abstract}
The theory of radiation-driven winds succeeded in describing terminal velocities and mass loss rates of massive stars. 
However, for A-type supergiants the standard m-CAK solution predicts values of mass loss and  terminal velocity 
higher than the observed values.
Based on the existence of a slow wind solution in fast rotating massive stars,  we explore numerically the parameter
space of radiation-driven flows  to search for new  wind solutions in slowly rotating stars, that could explain 
the origin of these discrepancies.
We solve the 1-D hydrodynamical equation of rotating radiation-driven winds at different stellar latitudes and explore the 
influence of ionization's changes throughout the wind in the velocity profile.
We have found that for particular sets of stellar and line-force parameters, a $new\, slow$ solution exists over the entire star 
when the rotational speed is slow or even zero. In the case of slow rotating A-type supergiant stars the presence of this novel slow
solution at all latitudes leads to mass losses and wind terminal velocities which are in agreement with the observed values.
The theoretical Wind Momentum-Luminosity Relationship derived with these slow solutions shows very good
agreement  with the empirical relationship. In addition, the ratio between the terminal and escape velocities, which provides a 
simple way to predict stellar wind energy and momentum input into the interstellar medium, is also properly traced.
\end{abstract}

%
%
\section{Introduction}

The theory of radiation driven winds or {\it CAK theory} \citep{cas75} and its later improvements m-CAK \citep{fri86,pau86} succeeded in
describing the terminal velocities ($V_{\infty}$) and mass-loss rates ($\dot{M}$) of very massive stars. Both CAK and m-CAK theories 
predict a tight relationship between the total mechanical momentum flow contained in the stellar wind outflow ($\dot{M}\,V_{\infty}$) and 
the stellar luminosity ($L$) of the mass-losing star, known as Wind Momentum-Luminosity (WM-L) relationship. 
The determination of a  WM-L relationship for A and B supergiants (Asgs and Bsgs) is important because it would allow the use of these 
stars as extragalactic distance indicators \citep{bre04}. 
This relationship had been first empirically found by \citet{kud95} for a sample of galactic O-B-A supergiants and giants. Its existence 
was confirmed for most luminous O-type stars by \citet{pul96}, who explained the difference of the WM-L relationship among Galaxy,
LMC and SMC in terms of their different abundances. Further observational studies of the WM-L relationship showed a strong dependence 
on the spectral type \citep{kud99} which was interpreted as an indication that the winds are driven by different sets of ions.

Although the CAK theory  has proved to be successful in explaining the global  mass--loss properties of O supergiants, the winds of Galactic 
mid-B supergiants are substantially weaker than predictions from the radiation-driven theory \citep{cro06}. Studies involving UV data \citep{pri05}
and radio observations \citep{ben07} have found discrepancies between empirical and predicted mass-loss rates. In most of the cases 
the supersonic regime of the wind is modeled with a velocity structure parametrized with a classical $\beta$--type law, with the $\beta$ 
exponent in the range 1--3, determined by fitting the H$\alpha$ line profile.

Similarly to mid- and late-Bsgs, the H$\alpha$ profile of Asg stars can be modeled with large $\beta$ values \citep{kud99}. In addition, 
the winds of Asgs show values of $V_{\infty}$ of about a factor of three lower than the predicted values \citep{ach97}. There is also observational 
evidence of a decrease of $V_{\infty}$ when increasing the effective escape velocity ${V}_{\mathrm{esc}}$, where 
${V}_{\mathrm{esc}}=\sqrt{2\, G\,M_{*}(1-\Gamma)\,/\,R_{*}}$ takes into account the effect of Thomson scattering on the gravitational 
potential through $\Gamma=\sigma_{\mathrm{e}}\,L_*/(4\,\pi\,c\, G\,M_{*})$, in clear contradiction with the standard radiation-driven wind theory 
\citep{ver98}. The existence of this negative slope was attributed to a change in the force multiplier parameter $\alpha$, either as a change 
in the ionization of the wind (via the parameter $\delta$) with distance or as a decoupling of the line driven ions in the wind from the ambient gas 
\citep{ach97}. The change in the ionization along the wind is often expressed by the difference between the parameters $\alpha$ and $\delta$; 
$\alpha_{\mathrm{eff}}\,=\,\alpha-\delta$ \citep{kud99}.

 When \citet{cur04} revisited the theory of steady rotating radiation driven winds, he obtained an exact formula for the location of the 
critical (singular) points and for the mass-loss rate. He showed that there exists another family of singular points, in addition to the 
standard m-CAK solution family (hereafter {\it fast wind solution}, FWS) when the star's rotation ($V$) is  close to the critical rotation 
velocity ($V_{\mathrm{crit}}$), that is  $\Omega\!= V/V_{\mathrm{crit}}\,\gtrsim$ 60 -- 70\%. 
The numerical solutions crossing through this other critical point family lead to winds with {\it lower} terminal velocities and {\it higher} 
densities ($\sim$ 30 times higher) than a non-rotating wind (FWS). He also found that for late B-type stars, 
these {\it slow wind solutions} (hereafter SWS) are also represented by a $\beta$--velocity law with $\beta\,>\,$1.

 Since the slow solutions might predict the formation of a circumstellar disk around fast rotating stars,  \citet{cur05} 
 modelled the density distribution of a rapidly rotating B[e] supergiant (with $V_{\mathrm rot}\,\sim\,$ 200 km s$^{-1}$; $\Omega \ga 0.6$) 
 assuming a change in the line-force parameters due to the bi-stability jump. This model leads to a fast wind in the polar regions and slow 
 outflows in the equatorial plane with density contrasts between the equator and the pole of about 10$^2$-10$^4$ near the stellar surface 
 ($r \la 2$ $R_{\ast}$) to values of 10$^1$-10$^2$ up to a  radii of $\sim $100 $R_{\ast}$.

However, none of the previously found solutions (FWS and SWS) are able to explain the observed velocities and mass losses in Asgs that, 
often, present low rotation speeds  ($V_{\mathrm rot}\,\sim\,$ 40 km s$^{-1}$; $\Omega <\,$ 0.4) and low outflow wind velocities. 
Nevertheless, we think that the large values of $\beta$ obtained empirically by \citet{kud99} for Asgs and Bsgs could be 
{\it related} to the presence of {\it{another type of}} slow wind solutions.

Therefore, based on \citet{ach97}'s hypothesis related to a change in the ionization of the wind, we investigate 
hydrodynamical solutions of rotating radiation winds for a wide range of line force multiplier parameters.  

In this work, we solve the 1-D hydrodynamical solution for rotating driven-winds and report the existence of {\it a new kind of solutions} 
obtained for slow rotating stars with high values of $\delta$, that resemble some of the properties of the SWS  found 
by  \citet{cur04} for fast rotating stars.

In  \S 2, we explore the influence of the ionization of the wind in the velocity profile of rotating radiation-driven flows, described by  the 
line--force parameter $\delta$. We find that for a particular set of line-force multiplier parameters $k$, $\alpha$ and $\delta$ in the 
range of the effective temperature of Asgs, there exists a  {\it new wind solution} that describes the properties of a weak 
outflow at all the stellar latitudes. This solution is obtained for {\it{low}} rotation rates, for instance 
$\Omega$ = $V/V_{\mathrm{crit}}\,$ =\,0.4, and even for the 
case without rotation, $\Omega\,=\,0$. Discussion of our results and conclusions are given in \S 3.

\section{Results}

We have solved numerically the 1-D hydrodynamic equations of rotating radiation-driven winds and obtained 
the radial velocity wind solutions as function of the latitude,  as described by \citet{cur04}.  We considered different rotational  
velocity rates, and different stellar and line force parameters. We adopt values for $\alpha$ = 0.49 -- 0.59 and $k$ = 0.37 -- 0.86 
that are in the range of  those computed by \citet{abb82,vin99} and \citet{shi94}, while the parameter  $\delta$ was arbitrarily 
selected  between 0.0 and 0.5, in order to study the influence of changes in ionization throughout the wind. This selection range 
allows us to find the new hydrodynamical solutions.

Table \ref{tab1} lists some of the models computed for a non-rotating Teff = 10\,000 K supergiant star with solar abundance. The 
first column indicates the model designation and  shows whether the solution is the new slow solution (s) or the fast solution (f).
Columns 2 to 5 quote a particular set of force multiplier parameters  ($\alpha$, $k$, $\delta$) and its corresponding 
$\alpha_{\mathrm{eff}}\,=\,\alpha-\delta$ value. For non-rotating A-type supergiants with low values of $\delta$ ($\lesssim\, 0.25$) we obtain 
the known 
m-CAK fast solutions at all latitudes, as expected. However, when $\delta \gtrsim 0.3$ the solution switches 
to a slow-acceleration mode. Regarding the velocity profile, this {\it new} slow solution is similar to the kind of slow solution 
reported by \citet{cur04} for fast rotating stars, but the mass-loss rate of this new solution, and therefore the wind density profile, 
is much lower than both, fast and slow (due to fast rotation) solutions. We were not able to  find any new slow wind solution in 
the interval 0.22 $<\,\delta\,<$ 0.30 for {\it all} latitudes.\\
Columns 6 and 7 list the corresponding terminal velocity and mass-loss rate. We can note that when increasing $\delta$ both 
the mass-loss rate and the terminal velocity decrease. The new solution yields $V_{\infty}\,<$\,$V_{\mathrm{esc}}$ (= 319 km\,s$^{-1}$), 
where $V_{\mathrm{esc}}$ was computed assuming a value of $\sigma_{\mathrm{e}}$\,= 0.33\,cm$^{2}$\,g$^{-1}$ \citep{ver98}. Finally, the last 
column quotes the wind modified momentum parameter, $D_{mom}$ = $\dot{M}\,V_{\infty}\, R_{\star}^{1/2}$ \citep{kud99}.

\begin{table*}[t!]
  \begin{center}
  \caption{{\small{Wind parameters for non-rotating stars ($\Omega$ = 0)
  with  $T_{\mathrm{eff}}$ = 10\,000 K, $\log\,g$ = 2, $R_{\star}$ = 60 $R_{\odot}$, $\log\,(L/L_{\odot})$ = 4.5 and $V_{\mathrm{esc}}$ = 319 km\,s$^{-1}$}.}}
  \label{tab1}
 {\scriptsize
  \begin{tabular}{lccrcccccccccc}
\hline
\hline
Model&$\alpha$&$k$&$\delta$&$\alpha_{\mathrm{eff}}$&$V_\infty$&$\dot{M}$&$D_{\mathrm{mom}}$\\
 &&&&&[km\,s$^{-1}$]&[M$_{\odot}$\,yr$^{-1}$]&[cgs]\\
\hline

W01 (f)& 0.49&  0.37&  0.00&  0.49&  546&    6.20\,$\times\,10^{-8}$ &   27.22\\
W02 (f)& 0.49&  0.37&  0.22&  0.27&  286&  6.48\,$\times\,10^{-9}$   &  25.96\\
W03 (s)& 0.49&  0.37&  0.30&  0.19&  201&  7.14\,$\times\,10^{-10}$&  24.85\\
W04 (f)& 0.49&  0.86&  0.00&  0.49&  552&  3.43\,$\times\,10^{-7}$   &   27.97\\
W05 (f)& 0.49&  0.86&  0.22&  0.23&  294&  1.43\,$\times\,10^{-7}$   &   27.31\\
W06 (s)& 0.49&  0.86&  0.42&  0.07&  174&  2.22\,$\times\,10^{-10}$&  24.28\\
W07 (f)& 0.59&  0.37&  0.00&  0.59&  786&  2.87\,$\times\,10^{-7}$  &   28.04\\
W08 (f)& 0.59&  0.37&  0.25&  0.34&  368&  1.03\,$\times\,10^{-7}$  &   27.27\\
W09 (s)& 0.59&  0.37&  0.34&  0.25&  242&  4.17\,$\times\,10^{-8}$ &   26.69\\
W10 (f)& 0.59&  0.86&  0.00&  0.59&  793&  1.19\,$\times\,10^{-6}$  &   28.66\\
W11 (f)& 0.59&  0.86&  0.26&  0.33&  365&  1.20\,$\times\,10^{-6}$  &   28.33\\
W12 (s)& 0.59&  0.86&  0.36&  0.23&  238&  1.14\,$\times\,10^{-6}$ &   28.12\\

\hline
  \end{tabular}
  }
 \end{center}
\end{table*}


\begin{table*}[t!]
  \begin{center}
 \caption{
 {\small{Slow wind solutions for rotating stars ($\Omega$ = 0.4).}}
  \label{tab2}}
\tabcolsep 4 pt
 {\scriptsize
  \begin{tabular}{lccrcccccccccccc}
\hline
\hline
Model&$T_{\mathrm{eff}}$&$\log\,g$&$R_{\star}$&$\alpha$&$k$&$\delta$&$\alpha_{\mathrm{eff}}$&$V_{\infty\,\mathrm{pol}}$&$F_{\mathrm{m\,pol}}$&$V_{\infty\,\mathrm{eq}}$&$F_{\mathrm{m\,eq}}$&$\dot{M}$&$\log\,(L/L_{\odot})$&$V_{\mathrm{esc}}$&$D_{\mathrm{mom}}$\\
&[kK]&&[$R_{\odot}]$&&&&&[km\,s$^{-1}$]&[M$_{\odot}$\,yr$^{-1}$\,st$^{-1}$]&[km\,s$^{-1}$]&[M$_{\odot}$\,yr$^{-1}$\,st$^{-1}$]&[M$_{\odot}$\,yr$^{-1}$]&&[km\,s$^{-1}$]&[cgs]\\\hline
R01 (s)& 11& 2& 70&  0.49& 0.37& 0.29& 0.20& 220& 6.99\,$\times\,10^{-10}$& 192& 9.09\,$\times\,10^{-10}$& 9.63\,$\times\,10^{-9}$ & 4.80& 286 & 26.03 \\
R02 (s)& 11& 2& 70&  0.49& 0.86& 0.33& 0.16& 209& 2.98\,$\times\,10^{-8}$ & 186& 4.18\,$\times\,10^{-8}$ & 4.21\,$\times\,10^{-7}$ & 4.80& 286 & 27.65 \\
R03 (s)& 11& 2& 70&  0.59& 0.37& 0.35& 0.14& 254& 1.59\,$\times\,10^{-8}$ & 226& 1.99\,$\times\,10^{-8}$ & 2.16\,$\times\,10^{-7}$ & 4.80& 286 & 27.44 \\
R04 (s)& 11& 2& 70&  0.59& 0.86& 0.36& 0.13& 255& 5.36\,$\times\,10^{-7}$ & 226& 6.82\,$\times\,10^{-7}$ & 7.33\,$\times\,10^{-6}$ & 4.80& 286 & 28.97 \\
R05 (s)& 11& 2& 60&  0.49& 0.86& 0.34& 0.15& 193& 2.39\,$\times\,10^{-8}$ & 173& 3.45\,$\times\,10^{-8}$ & 3.41\,$\times\,10^{-7}$ & 4.67& 317 & 27.49 \\
R06 (s)& 11& 2& 60& 0.59&  0.37& 0.35& 0.24& 237& 1.34\,$\times\,10^{-8}$ & 211& 1.67\,$\times\,10^{-8}$ & 1.82\,$\times\,10^{-7}$ & 4.67& 317 & 27.30 \\
R07 (s)& 10& 2& 60& 0.49& 0.37& 0.30& 0.19& 201& 5.92\,$\times\,10^{-11}$& 178& 7.78\,$\times\,10^{-11}$ & 8.16\,$\times\,10^{-10}$& 4.50& 308 & 24.88 \\
R08 (s)& 10& 2& 60& 0.49& 0.86& 0.33& 0.16& 194& 2.18\,$\times\,10^{-9}$& 173& 3.01\,$\times\,10^{-9}$   & 3.09\,$\times\,10^{-8} $& 4.50& 308 & 26.45 \\
R09 (s)& 10& 2& 60& 0.59& 0.37& 0.34& 0.25& 242& 3.32\,$\times\,10^{-9}$& 214& 4.11\,$\times\,10^{-9}$   & 4.50\,$\times\,10^{-8} $& 4.50& 308 & 26.70 \\
R10 (s)& 10& 2& 60& 0.59& 0.86& 0.36& 0.23& 238& 9.15\,$\times\,10^{-8}$& 212& 1.15\,$\times\,10^{-7}$   & 1.23\,$\times\,10^{-6} $& 4.50& 308 & 28.14 \\
R11 (s)& 10& 1.7& 80&  0.49& 0.37& 0.30& 0.19& 167& 1.40\,$\times\,10^{-9}$ & 148& 1.84\,$\times\,10^{-9}$ & 1.93\,$\times\,10^{-8}$ & 4.88& 259 & 26.24 \\
R12 (s)& 10& 1.7& 80&  0.49& 0.86& 0.34& 0.15& 161& 8.26\,$\times\,10^{-8}$ & 144& 1.21\,$\times\,10^{-7}$ & 1.18\,$\times\,10^{-6}$ & 4.88& 259 & 28.01 \\
R13 (s)& 10& 1.7& 80&  0.59& 0.37& 0.34& 0.25& 198& 3.01\,$\times\,10^{-8}$ & 176& 3.74\,$\times\,10^{-8}$ & 4.07\,$\times\,10^{-7}$ & 4.88& 259 & 27.64 \\
R14 (s)& 10& 1.7& 80&  0.59& 0.86& 0.36& 0.23& 198& 9.32\,$\times\,10^{-7}$ & 176& 1.20\,$\times\,10^{-6}$ & 1.28\,$\times\,10^{-5}$ & 4.88& 259 & 29.14 \\
R15 (s)& 9.5& 2& 60&  0.49& 0.37& 0.30& 0.19& 202& 2.26\,$\times\,10^{-11}$& 178& 2.94\,$\times\,10^{-11}$& 3.10\,$\times\,10^{-10}$& 4.41& 301 & 24.47 \\
R16 (s)& 9.5& 2& 60&  0.49& 0.86& 0.33& 0.16& 195& 7.14\,$\times\,10^{-10}$& 174& 9.95\,$\times\,10^{-10}$& 1.01\,$\times\,10^{-8}$ & 4.41& 301 & 25.96 \\
R17 (s)& 9.5& 2& 60&  0.59& 0.37& 0.34& 0.25& 242& 1.51\,$\times\,10^{-9}$ & 214& 1.86\,$\times\,10^{-9}$ & 2.04\,$\times\,10^{-8}$ & 4.41& 301 & 26.36 \\
R18 (s)& 9.5& 2& 60&  0.59& 0.86& 0.35& 0.24& 242& 4.06\,$\times\,10^{-8}$ & 213& 5.10\,$\times\,10^{-8}$ & 5.53\,$\times\,10^{-7}$ & 4.41& 301 & 27.79 \\
R19 (s)& 9.5& 1.7& 100&  0.49& 0.37& 0.30& 0.19& 184& 5.91\,$\times\,10^{-10}$& 163& 7.73\,$\times\,10^{-10}$& 8.14\,$\times\,10^{-9}$ & 4.86& 224 & 25.95 \\
R20 (s)& 9.5& 1.7& 100&  0.49& 0.86& 0.34& 0.15& 177& 2.53\,$\times\,10^{-8} $& 158& 3.60\,$\times\,10^{-8}$ & 3.59\,$\times\,10^{-7}$ & 4.86& 224 & 27.58 \\
R21 (s)& 9.5& 1.7& 100&  0.59& 0.37& 0.34& 0.25& 219& 1.70\,$\times\,10^{-8} $& 194& 2.11\,$\times\,10^{-8}$ & 2.30\,$\times\,10^{-7}$ & 4.86& 224 & 27.48 \\
R22 (s)& 9.5& 1.7& 100&  0.59& 0.86& 0.36& 0.23& 216& 4.90\,$\times\,10^{-7} $& 193& 6.21\,$\times\,10^{-7}$ & 6.68\,$\times\,10^{-6}$ & 4.86& 224 & 28.94 \\
R23 (s)& 9& 1.7&100&  0.49& 0.37& 0.33& 0.16& 175& 1.49\,$\times\,10^{-10}$& 157& 2.11\,$\times\,10^{-10}$& 2.10\,$\times\,10^{-9}$ & 4.76& 201 & 25.35 \\
R24 (s)& 9& 1.7& 100&  0.49& 0.86& 0.33& 0.16& 179& 2.77\,$\times\,10^{-8}$& 159& 3.89\,$\times\,10^{-8}$ & 3.91\,$\times\,10^{-7}$ & 4.76& 201 & 27.63 \\
R25 (s)& 9& 1.7& 100&  0.59& 0.37& 0.34& 0.25& 219& 1.70\,$\times\,10^{-8}$& 194& 2.11\,$\times\,10^{-8}$ & 2.30\,$\times\,10^{-7}$ & 4.76& 201 & 27.48 \\
R26 (s)& 9& 1.7& 100&  0.59& 0.86& 0.36& 0.23& 217& 4.90\,$\times\,10^{-7}$& 193& 6.21\,$\times\,10^{-7}$ & 6.67\,$\times\,10^{-6}$ & 4.76& 201 & 28.94 \\
\hline

  \end{tabular}
  }
 \end{center}
 \end{table*}

The differences {between the terminal velocities obtained} with the fast and slow solutions are remarkable (see Table \ref{tab1} column 6). 
These new slow solutions yield values of $\dot{M}$ in the range from 10$^{-6}$ to 10$^{-10}$ M$_{\odot}$\,yr$^{-1}$ and $V_{\infty}$ 
between 150 and 250 km s$^{-1}$. Thus, the ratio between fast and slow terminal velocities changes by factor of 3.0.

In the case of a slowly rotating A-type supergiants, we solve the 1-D hydrodynamical equations at all the latitudes and also found a family 
of slow wind solutions. In this paper,  we study solutions that show the same topology for all stellar latitudes, however we briefly want to 
mention a more general result related to the behaviour of the solutions in rotating radiation-driven winds. When $\delta$ is larger than 
a certain minimum value (typically in the range $0.2$ -- $0.25$) the wind solution switches from the fast solution to this {\it new} 
slow-acceleration mode at equatorial regions remaining the fast solution at higher latitudes up to the pole. 
The zones dominated by the slow solution enhance when increasing $\delta$ from this minimum value, and when it reaches a particular value of 
$\delta\,>\,\delta_{\mathrm{crit}}$  (e.g.: $\sim 0.3)$ this new slow solution {\it prevails at all latitudes}. \\
In a forthcoming paper, we will discuss in detail the behaviour of the different hydrodynamical solutions with spectral types.

Table \ref{tab2} lists some models for Asgs for which the fast and slow hydrodynamical solutions exist over the whole star. It quotes the 
$\delta_{\mathrm{crit}}$ value that  corresponds to the family of slow solutions at all latitudes.
Columns 1 to 4 indicate the model designation and the stellar parameters: effective temperature, stellar surface gravity and stellar radius, 
respectively.  Columns 5 to 12 quote the force multiplier parameters and the values of the terminal velocities and mass fluxes in  
polar ($V_{\infty \,\mathrm{pol}}$, $F_{m\,\mathrm{pol}}$) and equatorial ($V_{\infty \,\mathrm{eq}}$, $F_{m\,\mathrm{eq}}$) directions. Columns 13 to 16 list the total mass-loss rate, 
$\log\,(L/L_{\odot})$, $V_{\mathrm{esc}}$ and  $D_{\mathrm{mom}}$, respectively.

\begin{figure*}[ht]
\figurenum{1}
\epsscale{2.5}
\plottwo{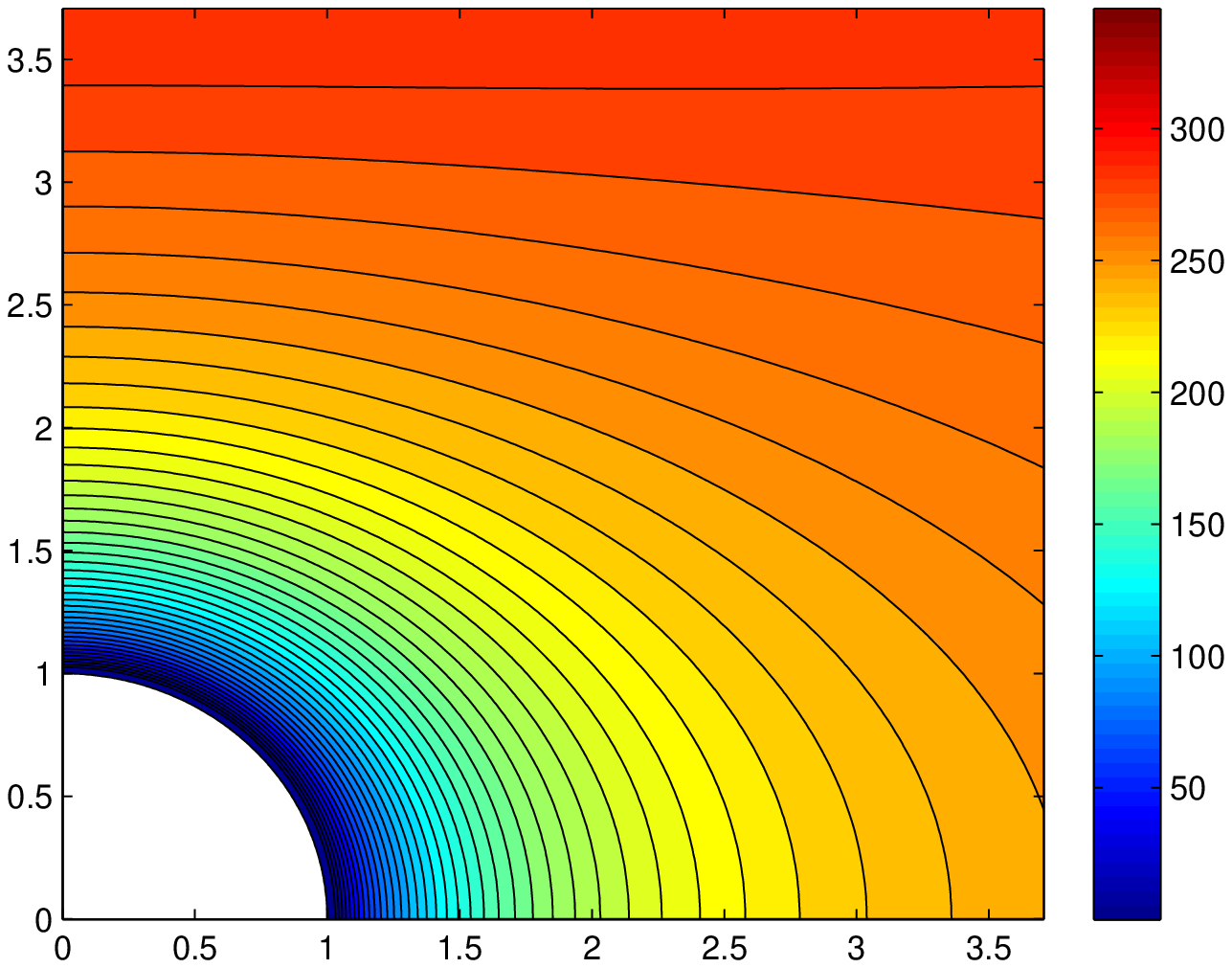}{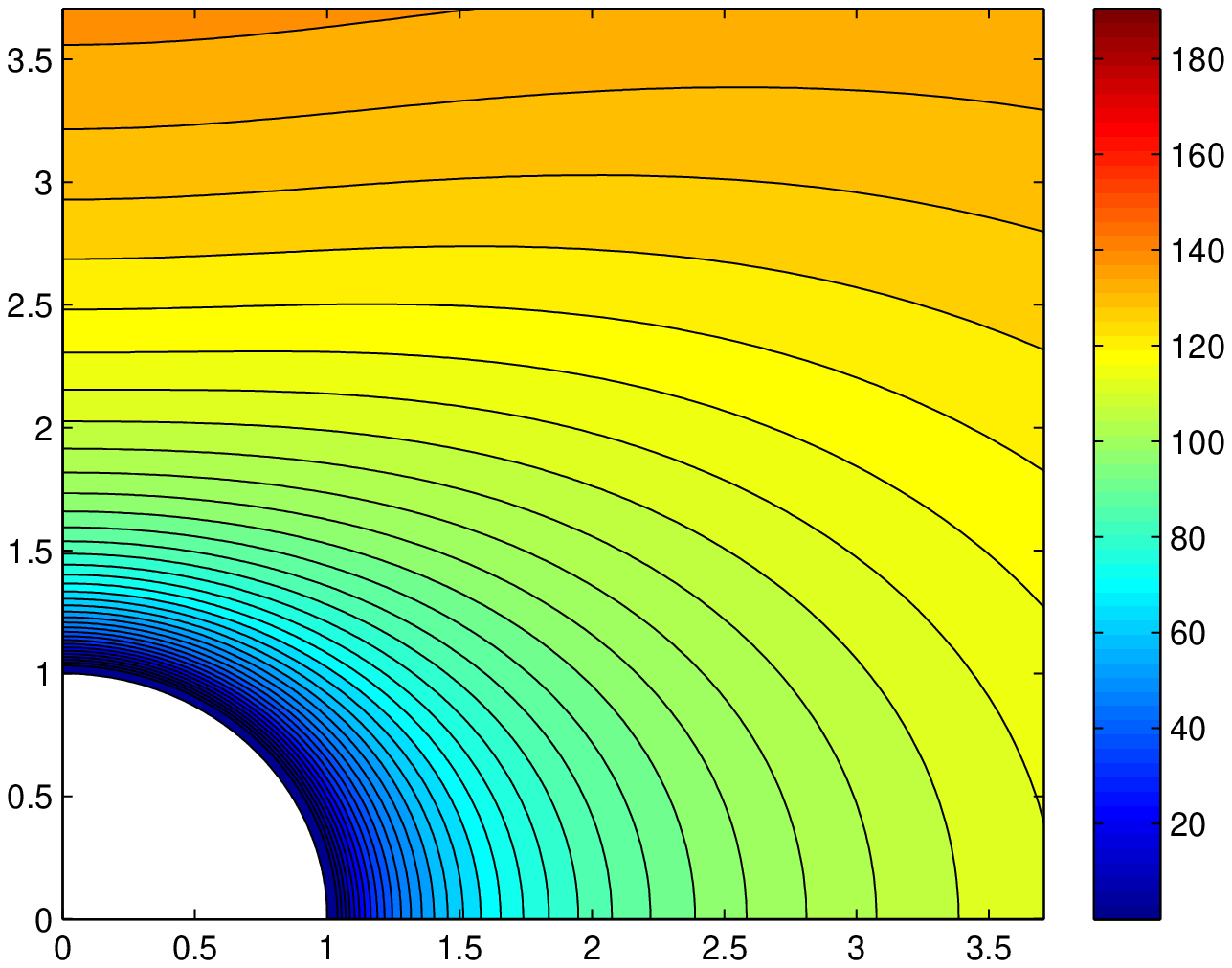}
\caption{
{\small{Latitude dependence of the wind velocity distribution for fast (left panel) and slow (right panel) solutions. 
The parameters of the solution correspond to model R08, except for the different values of $\delta$; fast solution is with $\delta\,=\,0.15$. 
The units of the axes are stellar radii and the abscissa defines the equatorial direction. The scale bars indicate the radial velocity in km s$^{-1}$.}}
 \label{fig1}}
\end{figure*}

Figure \ref{fig1} displays the behaviour of the fast and slow solutions as function of stellar latitude for a rotating star with $\Omega$ = 0.4. 
The models share the same stellar and wind parameters  except for  $\delta$: fast solution was computed with $\delta$\,=\,0.15 
and slow solution with $\delta$\,=\,0.30 (we chose as example model R08 from Table \ref{tab2}).

\begin{figure}[ht]
\figurenum{2}
\includegraphics[width=2.8in]{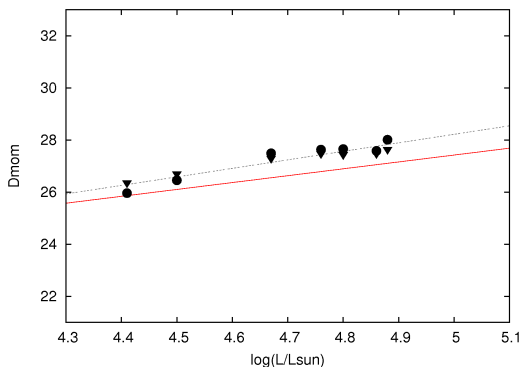}
\caption{
{\small{The WM-L relationship derived from theoretical data computed  from new slow wind models with $\Omega=0.4$ 
and the following sets of parameters, $\alpha\,=0.59$; $k\,=0.37$ (black downward triangles) and $\alpha\,=0.49$; $k\,=0.86$ (black circles), 
see Table \ref{tab2}.  The theoretical WM-L relationship (dashed line) shows a good agreement with the  observational relationship (red solid line) taken 
from \citet{kud99}. \label{WML}}}}
\end{figure}

Finally, we want to remark  that the theoretical wind parameters ($\dot{M}$, $V_{\infty}$) related to this new slow wind solutions are 
in agreement with those observed in Asg stars \citep{ach97,ver98,ver99,kud99}. Moreover, the computed values for $D_{\mathrm{mom}}$ using 
these new slow solutions follow the  trend of the observed WM-L relationship for Asgs (as it is shown in Figure \ref{WML}). A good agreement 
was obtained with the values reported by \citet{abb82}, $\alpha\,=\,0.59$, $k\,=\,0.37$, however models with different set of line--force 
parameters, i.e $\alpha\,=\,0.49$, $k\,=\,0.86$, fit, as well.

\begin{figure}[ht]
\figurenum{3}
\includegraphics[width=2.8in]{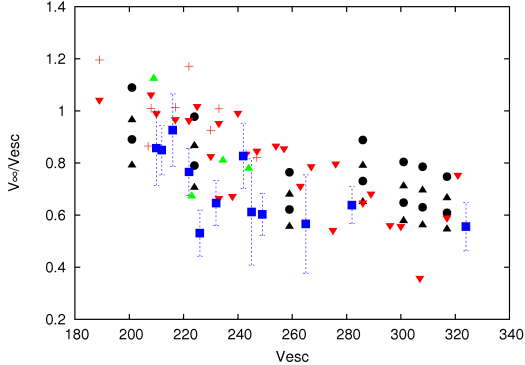}
\caption{Relation between $V_{\infty}$/$V_{\mathrm{esc}}$ vs. $V_{\mathrm{esc}}$ corresponding to polar (black circles) 
and equatorial (black triangles) slow solutions. Down-triangles and  crosses (red symbols) represent the observational 
data taken from \citet{ver98}; the crosses indicate terminal velocities obtained from saturated PCygni UV lines whereas 
the down-triangles correspond to values determined by means of discrete absorption components; up-triangles (green) 
correspond to terminal velocities from \citet{kud99}; squares (blue) represent the measurements provided by \citet{ach97} 
with their  error estimates. Slow wind solution follows the same trend of the observations. \label{trend}}
\end{figure}

We also find that  this new theoretical slow wind solutions predict a decreasing relation of $V_{\infty}$/$V_{\mathrm{esc}}$ with respect to 
$V_{\mathrm{esc}}$  with the same observational trend found by \citet{ver98}. Figure \ref{trend} shows our numerical results together with 
observational data obtained from the literature. The down triangles and crosses (red symbols) represent the observational data of \citet{ver99};
the crosses indicate terminal velocities obtained from saturated PCygni UV lines whereas the triangles represent values obtained  
by means of discrete absorption components; the green up-triangles correspond to terminal velocities from \citet{kud99}; the blue 
squares represent the measurements provided by \citet{ach97} with their  error estimates. Our slow solution results are plotted 
in black symbols (circles/triangles for polar/equatorial directions) and nicely follow the same trend.

%

\section{Discussion and Conclusions}

Previous studies on radiation-driven winds based on 1-D high-rotating early-type stars ($\Omega\!>\!$ 60-70\%) carried out by  \citet{cur04} 
demonstrated  the existence of slow wind solutions. These types of solutions predict higher mass-loss rates and lower 
flow speeds at the equatorial plane than those of the polar zones and, as a consequence, a disk-like structure can be 
formed. In the present work, we explored numerically the parameter
space of radiation-driven flows, particularly for high values of $\delta$,  and found a new kind of  wind solutions 
for slowly rotating stars ($\Omega\!<\!$ 40\% and even without rotation). In order to distinguish the difference between 
the new solutions reported here and those obtained by \citet{cur04}, we propose to call them  {\textit{``low-$\Omega$,\,
high-$\delta$ SWS''}} and {\textit{``high-$\Omega$,\,low-$\delta$ SWS''}}, respectively.

 Although the {\textit{low-$\Omega$,\,high-$\delta$ SWS} were computed using  1-D hydrodynamic equations, they 
 provide a complete understanding of the dynamical outflow in slow rotating stars and settle solid basis for the computation 
 of multidimensional hydrodynamic models. In our particular case, the 1-D approximation is a good approach to describe the 
 slow wind properties of Asgs since the deformation of the star due to a slowly rotational speed can be neglected. On the other 
 hand, in a low rotating star the density contrast between the equator and the pole is very low and,  therefore, the wind would 
 present a quasi-spherical distribution.}

We found that {\textit{low-$\Omega$,\,high-$\delta$ SWS}} properly trace the ratio between the terminal and escape 
velocities, which provides a simple way to predict stellar 
wind energy and momentum input into the interstellar medium. In addition, the new solutions  follow the observational 
trend of $V_{\infty}$/$V_{\mathrm{esc}}$\,vs.\,$V_{\mathrm{esc}}$ reported by \citet{ver98}. Our results support \citet{ver98} 
hypothesis stating that the negative slope of the latter relation could be linked to the degree of ionization and the density of 
the wind.\\
Moreover, the theoretical WM-L relationship derived with \textit{low-$\Omega$,\,high-$\delta$ SWS} shows a good 
agreement with the empirical relationship and brings back to the idea of using these stars as extragalactic distance indicators 
\citep{bre04}.
 
Taking into account the previous results, we think that {\textit{low-$\Omega$,\,high-$\delta$ SWS}} might  help to understand 
the long-standing problem of {\it weak winds} \citep[see e.g.][]{pul08} because these solutions predict, besides slower 
terminal velocities, values of mass-loss rates that might be some hundreds times lower than the standard or fast solutions, 
which precisely  corresponds to the observed discrepancy between theory and observations.\\
The advantage of the model is that the new results stand on a radiation-driven wind characterized by the line-force parameters. 
Although the solution was obtained using ad-hoc values of the parameter $\delta$, the  calculation of this parameter should 
be revisited in order to introduce the effects of the variation of the ionization of the wind with distance. Future experiments 
to search for high values of the parameter $\delta$ in Asgs could be performed fitting line spectral features of different degree 
of ionization with synthetic line profiles computed with the radiative transfer equation and the new hydrodynamical wind model.

\begin{acknowledgements}

We thank the Referee for his/her helpful comments on our manuscript. MC acknowledges financial support from 
Centro de Astrof\'{\i}sica de Valpara\'{\i}so. LC acknowledges financial support from the 
Agencia de Promoci\'on Cient\'{\i}fica y Tecnol\'ogica (BID 1728 OC/AR PICT 111), from CONICET (PIP 0300), and the Programa 
de Incentivos G11/089 of the Universidad Nacional de La Plata, Argentina.
\end{acknowledgements}
\bibliographystyle{apj}
\bibliography{citas}

\end{document}